# A Magnetically and Electrically Powered Hybrid Micromotor in Conductive Solutions: Synergistic Propulsion Effects and Label-Free Cargo Transport and Sensing

Yue Wu[1], Sivan Yakov[2], Afu Fu[3] and Gilad Yossifon[1,2]*

[1]School of Mechanical Engineering, University of Tel-Aviv, Tel-Aviv 69978, Israel

[2]Faculty of Mechanical Engineering, Micro- and Nanofluidics Laboratory, Technion – Israel Institute of Technology, Haifa 32000, Israel

[3]Technion Integrated Cancer Center, The Rappaport Faculty of Medicine and Research Institute, Technion-Israel Institute of Technology, Haifa 3109602, Israel* Corresponding author: gyossifon@tauex.tau.ac.il

## Abstract

Electrically powered micro- and nanomotors are promising tools for in-vitro single-cell analysis. In particular, single cells can be trapped, transported and electroporated by a Janus particle (JP) using an externally applied electric field. However, while dielectrophoretic (DEP)-based cargo manipulation can be achieved at high-solution conductivity, electrical propulsion of these micromotors becomes ineffective at solution conductivities exceeding ~0.3mS/cm. Here, we successfully extended JP cargo manipulation and transport capabilities to conductive near-physiological (<6mS/cm) solutions by combining magnetic field-based micromotor propulsion and navigation with DEP-based manipulation of various synthetic and biological cargos. Combination of a rotating magnetic field and electric field resulted in enhanced micromotor mobility and steering control through tuning of the electric field frequency. In addition, we demonstrated the micromotor's ability of identifying apoptotic cell among viable and necrotic cells based their dielectrophoretic difference, thus, enabling to analyze the apoptotic status in the single cell samples for drug discovery, cell therapeutics and immunotherapy. We also demonstrated the ability to trap and transport live cells towards regions containing doxorubicin-loaded liposomes. This hybrid micromotor approach for label-free trapping, transporting and sensing of selected cells within conductive solutions, opens new opportunities in drug delivery and single cell analysis, where close-to-physiological media conditions are necessary.

## 1. Introduction

Hybrid micromotors, which combine different propulsion mechanisms (e.g., electric and thermal[1], microorganism and magnetic[2], acoustic and light[3], chemical and magnetic[4]), have shown great potential for biomedical applications due to their enhanced robustness of operation and motion control. Among these mechanisms, electrically powered micromotors offer several advantages, such as unified and selective cargo loading, transport and cargo release[5,6,7], localized electroporation[8] and localized electro-deformation of cell/nucleus as a mechanical biomarker[7]. However, motion of micromotors, e.g. metallo-dielectric Janus particles (JPs), actuated by an AC electric field is quenched with increasing solution conductivity[9] and practically ceases at solution conductivities exceeding[10,11] ~0.3mS/cm. In order to manipulate JPs and still maintain cargo manipulation and localized electroporation capabilities offered by the electric field, in high conductivity solutions, it is essential to propel the JP using other methods, such as thermophoresis[12], optics[13], chemical[14,] and catalytic reaction[15], or magnetism[16].

In biological systems, the solutions required to ensure prolonged survival are highly conductive. For example, the conductivity of human blood serum is 10-20mS/cm[17], mammalian cell culture medium is 10-20mS/cm and phosphate-buffered saline (PBS) is 13mS/cm. Although the JP cannot be electrically propelled in highly conductive media[9], dielectrophoretic force on biological organisms in such solutions is still present. For example, the dielectrophoretic behavior of bacteria was studied in solution conductivities ranging between $1\times10^{-4}$mS/cm and 1S/cm[18]. Different micromotor propulsion mechanisms have been studied in solutions of high conductivity. Magnesium reaction-based micromotors were designed for drug delivery in gastric fluid (pH ~1.3, conductivity >10mS/cm)[19]. Micromotors made of photoactive (CdTe) and photocatalytic ($Fe_3O_4$) parts were designed to propel in glucose media (<10μS/cm) and human blood serum (10-20mS/cm) upon the application of visible light[20].

Micromotor actuation in biological samples by application of a magnetic field has become attractive because of its applicability to biological medium[21], ability to move in the bulk without relying on its interaction with boundaries[22], no requirement for fuel or direct contact between the external magnet and tissues[23], steering accuracy[24], and the ability to operate in a wide range of temperatures and solution conductivities (even in vivo[25]). Magnetic microparticles actuated in fluids, using either an external uniform rotating magnetic field[26] or magnetic field gradients (i.e. magnetophoresis[21], have enabled applications, such as magnetic cell carriers[16,27,28], probes in biophysical studies[29], imaging-guided therapy[23], and micro-injections for drug delivery[30,31]. Spherical magnetic particles were shown to propel under rotating uniform magnetic field due to their rolling over surfaces[26]. A peanut shaped micromotor [32] has been shown to move in both rolling and wobbling modes to enable climbing over steep slopes, while a rotating spiral micromotor[33] has been shown to have the potential to for in-vivo non-invasive Zygote transfer, and a rolling and tumbling magnetic biohybrid micromotors for drug delivery in the small intestine of a mouse[34]. Magnetic field actuation of micromotors is commonly combined with an additional mode of actuation to form hybrid micromotors, as in the case of combined magnetic and electric fields to assemble mobile micromachines[35], magnetic steering of catalytically driven micromotors within sea water[36] as well as targeted drug delivery using sperm cells with a magnetic cap[2,37]. Herein, we present a novel combination of rotating magnetic fields[38] and electric fields for sustained micromotor propulsion as well as manipulation of label-free dielectrophoresis (DEP)-based organic and inorganic cargo in conductive near-physiological solutions. In addition, novel synergistic effect of combining magnetic rolling and electric field-driven propulsion for enhanced micromotor mobility and steering control is studied. These are then complemented with two biologically relevant potential micromotor-based applications of trapping and transporting live cells towards regions containing doxorubicin-loaded liposomes as well as identifying between apoptotic, necrotic and viable cells based on the difference in their DEP characteristics.

## 2. Results

### 2.1. Design and characterization of the hybrid micromotor system

The metallo-dielectric JP micromotor was fabricated by coating green fluorescent polystyrene spheres with 15nm Cr (for adhesion[39]), and then with 50nm Ni (ferromagnetic material for magnetization) and 15nm Au (as a highly conductive inert layer), as described in our previous studies[6]. It is noted that all deposited layers (Cr, Ni, and Au) are conductive materials (Cr: $7.9\times10^6$S/m, Ni: $1.4\times10^7$S/m, and Au: $4.2\times10^7$S/m at room temperature) that contribute to the increased conductivity of the metallic coated JP's hemisphere. The micromotor was magnetized between two magnets, with their magnetic field aligned parallel to the metallo-dielectric interface of the JP, as illustrated in Fig.1A (see Fig. S1 for the simulated magnetic vector field used for magnetization of the ferromagnetic 50nm Ni layer in both current and previously studied[7,8] configurations with the magnetic field applied tangential and normal to the metallo-dielectric interface, respectively). The micromotors were propelled and steered using an external magnetic field, through an experimental apparatus consisting of two motors, a magnet and a gliding track system, onto which the rotating magnet was fixed (Fig.1B). An external rotating magnetic field (Fig.1B), induced by a fixed block magnet (neodymium magnet, grade:N35), rotated the JP along the axis perpendicular to its metallo-dielectric interface (*x* axis in Figs.1B-D) (see Fig. S2 for the simulated magnetic field vector and gradient that are induced by the block magnet). The setup enabled precise micromotor steering along any two-dimensional trajectory, e.g. square (Fig.1E) or triangle (Fig.1F), by controlling the location of the magnet along the gliding track (Fig. 1B, Fig.1E-F and Video 1 in Supplementary material). Increasing the solution conductivity by increasing the KCl concentration in the solution (see Methods), increased the magnetic rolling velocity of the JPs (Fig.1G), likely due to the decreased electric double layer (EDL) with increased conductivity, leading to increased friction of the JP with the bottom substrate on which it rolls. Demirors et al[40] has recently shown that intensification of an applied electric field can also lead to a decreased (increased) distance between a rolling JP and a bare (SiO$_2$ coated) indium tin oxide (ITO) conducting substrate and, with it, to increased (decreased) friction, which, in turn, resulted in increased (decreased) JP rolling velocity. In addition, the magnetic and electric fields had a synergetic effect on the overall mobility of the JPs, as shown in Fig.1H and elaborated on in the following section.

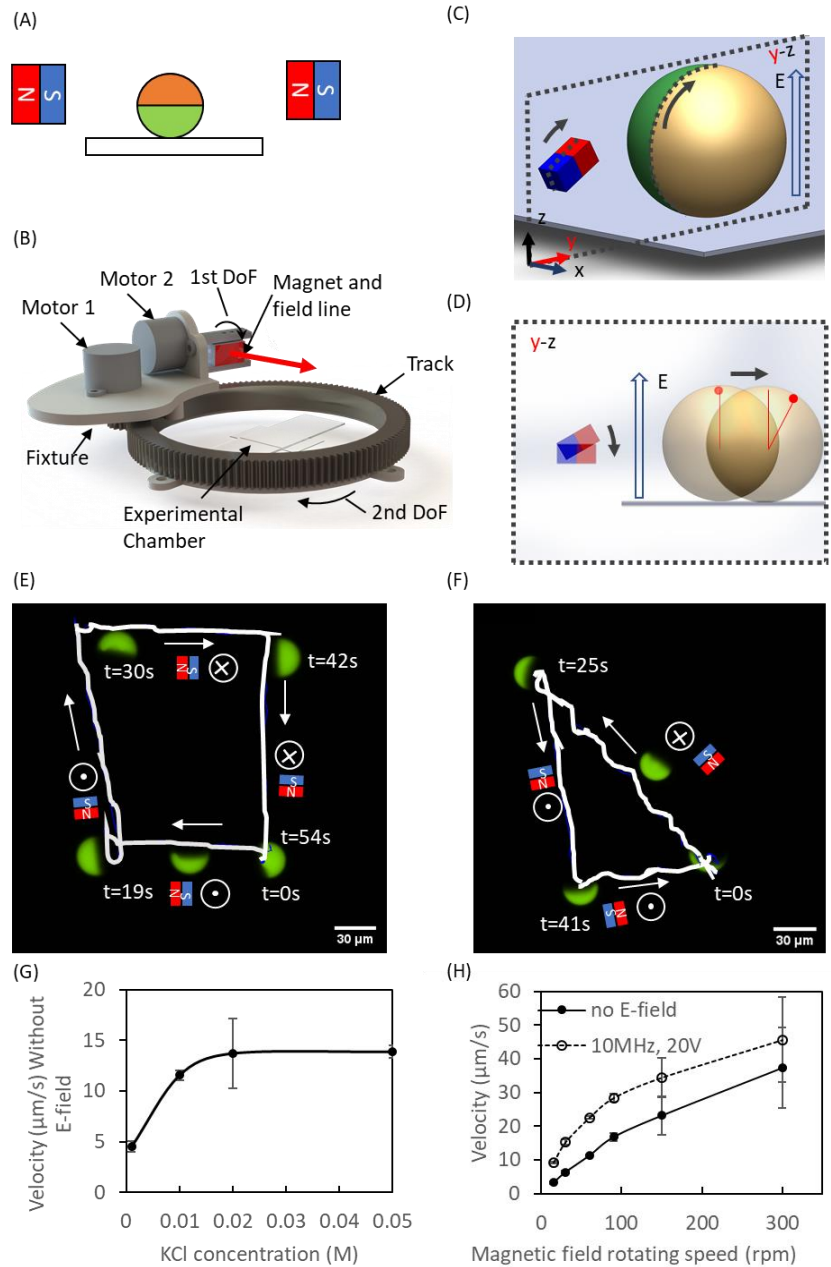

**Figure 1. JP propulsion and steering using a rotating magnetic field, in combination with an electric field for alignment of the JP's metallo-dielectric interface.** (A) Magnetization of the Janus particle (JP). (B) Schematic of the experimental apparatus. The experimental chamber is placed 40 mm away from the magnet. The magnet and motor are all mounted on a fixture that is mounted on a track. Schematic (C) isometric and (D) side view of the JP propulsion mechanism. Magnetic steering of the JP along (E) square- or (F) triangle-shaped trajectories under an externally applied electric field of 1MHz, 20V, and within a 0.02M KCl solution (conductivity: 2.2-2.6mS/cm, pH: 6.14) (see Video 1 in Supplementary materials). (G) Enhanced translational velocity of rotating magnetic field-driven JPs (27μm) due to increased KCl concentration. (H) Enhanced translational velocity of rotating magnetic field-driven JPs (27μm) upon electric field-assisted (10MHz, 20V) alignment of their metallo-dielectric interface within a 0.01M KCl solution (1-1.3mS/cm, pH: 6.14).

## 2.2. Influence of the electro-orientation of the metallo-dielectric JP interface on the magnetic rolling velocity

With application of an electric field normal to the substrate the metallo-dielectric interface orients approximately parallel to the applied electric field due to the electro-orientation torque, which is the cross-product of the induced dipole ($\theta_E$) and the applied field (E). On further application of an external rotating magnetic field, the JP's metallo-dieletric interface oscillated around its putative orthogonal orientation relative to the bottom substrate, as shown in Fig.2A (see Video 2 in the Supplementary materials). These oscillations can be characterized using an oscillation factor defined as $A_2/A_1$, where $A_2$ is the area of the bare polystyrene surface and $A_1$ is the area of the entire JP, which can be plotted against time for various solution conductivities (Fig.2B). These oscillations can also be characterized using an oscillation factor range defined as $D_1-D_2$, where $D_1$ and $D_2$ represent the maximum and minimum values of the oscillation factors, respectively, as shown in Fig.2C. It was found that $D_1-D_2$ increased with solution conductivity and, at solution conductivity of 0.05M KCl (6mS/cm, pH: 6.14), it almost approached the behavior of that measured without an applied electric field with $D_1-D_2$ approaching unity.

The numerically calculated electric fields depicted in Fig.2F-G provide an explanation as to why JP oscillation increased with conductivity. As solution conductivity increased, so does the characteristic frequency of the induced-charge electric double layer (EDL). The latter is the inverse of the resistor-capacitor (RC) time of the induced-charge EDL. Hence, for any given applied electric field frequency (e.g., 50kHz, as depicted in Fig.2F-G, the screening of the metallic coating of the JP by the induced-charge double-layer becomes more effective with increased solution conductivity and the electric field streamlines become more tangential to the metallic coating (Fig.2G). In the case of the lower-conductivity solution, the applied frequency is much higher than the RC frequency, such that the induced-charge EDL does not have time to form and there is practically no screening of the metallic coating, with the electric field becoming normal to the metallic coating (Fig.2F). As a result, with increasing conductivity, the electrostatic solution of the electric potential and field induced on the metallic side approaches that of the dielectric side, such that in the limit of very high conductivity, they become identical and effectively that of an insulated particle. This leads to the loss of the dominant dipole induced within the metallic coating in the non-screened (i.e. negligible induced EDL) case, thus, leading to a reduced electro-orientation torque. From the above experimental results (Fig.2C-D), a solution conductivity of ~0.05M KCl (~6mS/cm) can be considered as the upper limit for maintaining stable magnetic rolling. While applying electric fields at frequencies exceeding the RC frequency, the vanishing of the induced EDL that screens the metallic coated hemisphere stabilizes the metallo-dielectric interface, resulting in increased magnetic rolling mobility. Low-frequency electric fields result in an opposite effect, wherein, the magnetic rolling velocity decreases below that obtained without an electric field (Fig. 2E). A possible explanation for this loss of mobility is that the alternating current electro-osmotic (ACEO) flow[41] induced at these low frequencies leads to a levitation force on the JP, which, in turn, results in decreased friction with the bottom substrate.

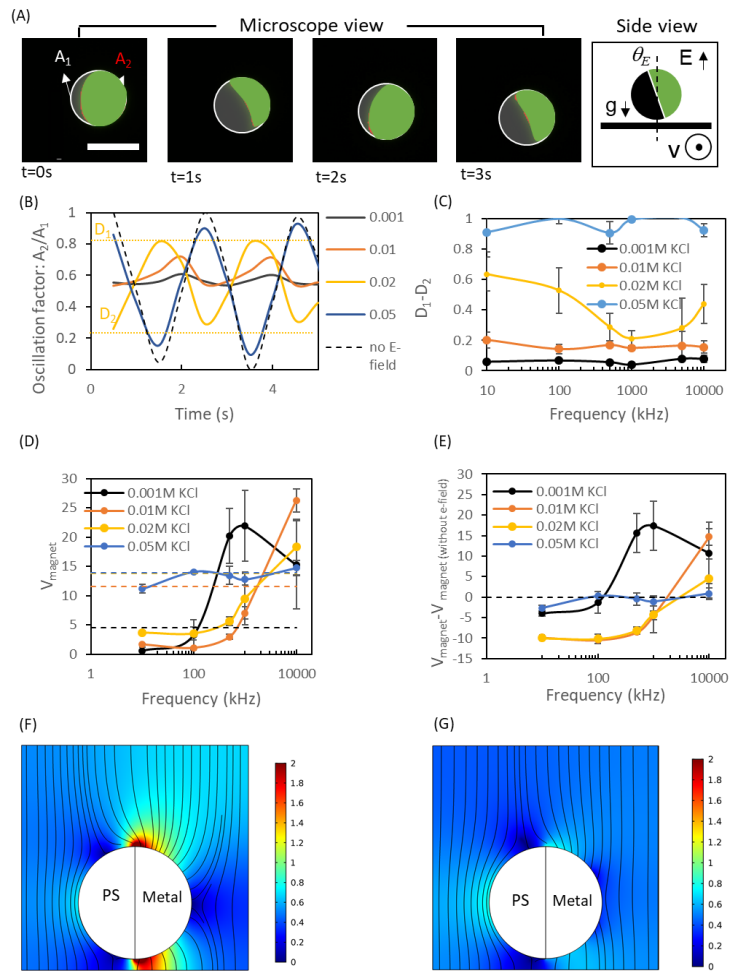

**Figure. 2. Synergetic effect of the electric field and rotating magnetic field on JP (27μm) velocity** (A) Time-lapse of the JP metallo-dielectric interface orientation under an applied field of 100kHz, 20V, within 0.02M KCl solution (2.2-2.6mS/cm, pH:6.14), and an external magnetic field rotation speed of 30rpm. The total cross-section area at the equator of the JP is marked with a white circle ($A_1$) and the area of the polystyrene part is marked in green ($A_2$). The black color is the gold. The oscillation factor is defined as $A_2/A_1$. See Video 2 in Supplementary materials. Scale bar: 30μm. (B) The oscillation factor over time for varying solution conductivities, under an applied field of 100kHz, 20V. The maximum and minimum values of the oscillation factor for 0.02M KCl are marked as $D_1$ and $D_2$, respectively. (C) $D_1-D_2$ plotted versus frequency for varying solution conductivities. (D) $V_{magnet}$ of the JP velocity versus frequency for varying solution conductivities. The dashed lines represent the no electric field case of each of the corresponding conductivities, as shown in Fig.1G. (E) $V_{magnet}-V_{magnet\ (without\ E-field)}$ of the JP velocity versus frequency for varying solution conductivities. (F-G) Numerical simulation of the electric field streamlines and intensity for (F) 0.001M KCl (0.1mS/cm), 50kHz, and (G) 0.02M KCl (2.6mS/cm), 50kHz.

## 2.3. Influence of magnetic rolling on the electric field-based JP translational velocity

From our[9] and others[10] previous studies, both induced-charge electro-phoretic (ICEP[42,43,44]) and self-dielectrophoretic (sDEP) velocities of JPs are reduced in high-conductivity solution ($7\times10^{-5}$M to $5\times10^{-4}$M KCl). In the current study, it was observed that without an external rotating magnetic field, the electrical propulsion of the JP, $V_{electric}$, in moderate-conductivity solutions was almost zero (Fig.3B, 3s-10s). On the other hand, the trajectory of a JP due to only rotating magnetic field, without applied electric field, is orthogonal to the axis of magnet rotation (Fig.3A). However, upon a combined application of an electric field with an external magnetic field rotation, the electric field-driven propulsion, $V_{electric}$, which is orthogonal to the magnetic rolling-induced velocity, became observable (Fig.3B, 10s-30s) in a 0.01M KCl solution, under 1MHz, 20V. Moreover, in a relatively lower-conductivity solution, JP movement could be manipulated by both ICEP and sDEP propulsion modes (Fig.3C) that are controlled by the applied electric field frequency (Fig.3D). $V_{electric}$ almost vanished in solution concentrations above ~0.02M KCl (Fig.3E), regardless of the field frequency (1kHz-10MHz).

The speed of the external magnetic field rotation also affected $V_{electric}$, as shown in Fig.3F, wherein the $V_{electric}$ initially increased when the external magnetic field rotation speed increased. However, beyond a certain rotation velocity, the JP translating velocity reached approximate saturation. There are several mechanisms that may contribute to a lift force that results in a decreased friction of the JP with the substrate, thereby enabling an increased $V_{electric}$ of the JP. The Magnus effect[45] may generate a lift force on a rotating and translating sphere (rolling motion is a combination of rotation and translation) by ignoring the flow asymmetry due to wall proximity effects[46,47]. A order of magnitude approximation of the magnus force $F_{magnus}=\pi\rho R^3\omega u$, where $\rho$ is the fluid density (1000kg/m$^3$), $R$ is the radius of JP (13.5μm), $\omega$ is the angular speed (31rad/s), $u$ is the relative linear velocity between the particle and the fluid (at full contact $\omega R$=424μm/s), the magnus force is $1.0\times10^{-13}$N, which is comparable to the immersed weight ($4.8\times10^{-12}$N).

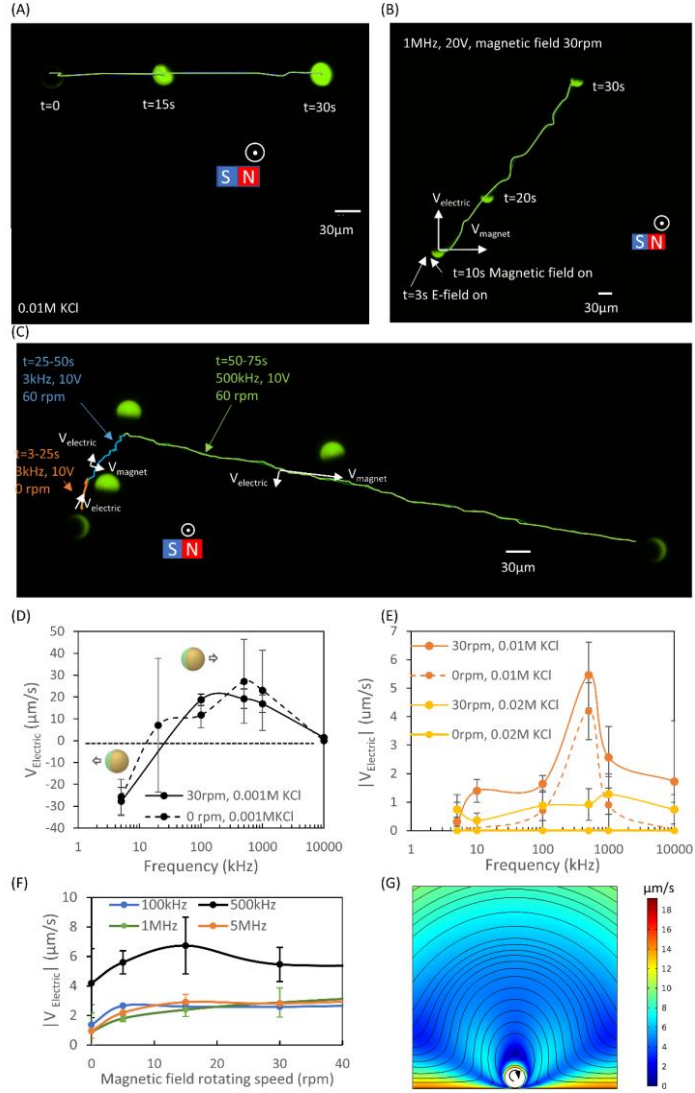

**Figure. 3. The impact of magnetic rolling of the JP on the electric field velocity ($V_{electric}$) in conductive solution.** (A) The trajectory of a JP propelled by magnetic field rolling without an electric field, within 0.01M KCl (1-1.3mS/cm, pH: 6.14) solution; only the $V_{magnet}$ component is apparent (See Video 3 in Supplementary materials). (B) The trajectory of a JP propelled by both magnetic field rolling and an electric field (1MHz, 20V) in moderate-conductivity solution (0.01M KCl). The resultant velocity is the combination, i.e. vector sum, of both $V_{magnet}$ and $V_{electric}$. (see Video 3 in Supplementary materials). (C) A time-lapse image of the JP demonstrating the combined effect of magnetic field propulsion and electric field propulsion in both ICEP (3kHz, 10V) and sDEP (500kHz, 10V) modes within 0.001M KCl (conductivity: 0.1-0.15mS/cm, pH: 6.14). (see Video 3 in Supplementary materials). (D-E) $V_{electric}$ versus frequency for a JP within solutions of varying conductivities and an applied voltage of 20V, with/without an externally rotating magnetic field. D: 0.001M KCl; E: 0.01M and 0.02M KCl (conductivity: 1-1.3mS/cm, pH: 6.14). (F) $V_{electric}$ versus external magnetic field rotating speed for a 27μm JP in 0.01M KCl at 20V, for various frequencies. (G) Simulation of velocity field around a rolling sphere. The boundary conditions are listed in Supplementary material Table S1.

## 2.4. Selective cargo trapping and transport

The micromotor system was able to trap and transport particles (Fig.4A) exhibiting nDEP response at the equator of the metallic-coated hemisphere (location 1 in Fig.4B), where an electric field minimum exists. Particles that exhibited pDEP were trapped at the equator of the polystyrene hemisphere (location 2 in Fig.4B), where an electric field maximum exists. Although there are additional pDEP and nDEP trap locations corresponding to local electric field maximum and minimum, respectively (Fig.4B, e.g. pDEP between the metallic coating and the bottom substrate and at the top of the metallic coating, as well as nDEP traps on the top and bottom parts of the dielectric side), the magnetic rotation induced very strong hydrodynamic shearing forces at these locations (Fig.4D). The minimal hydrodynamic shear at the poles of the metallic and dielectric hemispheres (Fig.4D C-C), resulted in effective nDEP and pDEP trapping of cargo, respectively. The DEP force can be expressed as $F_{DEP} = \pi r^3 \varepsilon_m Re[K(\omega)] \nabla E^2$, where $r$ is the particle radius, $\varepsilon_m$ is the permittivity of the suspending medium, $\nabla$ is the del vector operator, $E$ is the electric field norm and $Re[K(\omega)]$ is the real part of the Clausius-Mossotti factor[48]. This relation shows that the DEP force linearly scales with the volume of the trapped cargo. For the same applied field, it also suggests that it decreases with increased JP size as it linearly scales with the gradient of $E^2$ (see also simulations of the DEP force distribution in Fig. S3 for varying JP size indicating the trap's potential becoming deeper and wider with increased JP size). Hence, since increased JP size enhances both its cargo loading capability and its rolling velocity (scales linearly with the JP diameter), we chose to use a relatively large JP of 27µm in diameter.

Polystyrene cargos (10µm) were used to study the JP trapping capacity (i.e., maximum number of trapped particles) in various solution conductivities and under external magnetic rotation speeds (see Video 4 in Supplementary materials). Fig.4C illustrates the time-lapse images of a 27µm micromotor trapping three polystyrene particles under 100kHz, 20V, and 30rpm within a 0.01M KCl solution. As shown in Fig.4E, the maximum number of trapped cargos decreased with increasing solution conductivity due to decreased nDEP force (see Fig.2 (F-G)). Also, as shown in Fig.4F, the number of trapped particles decreased with increasing external magnet rotation speed due to increased hydrodynamic shear.

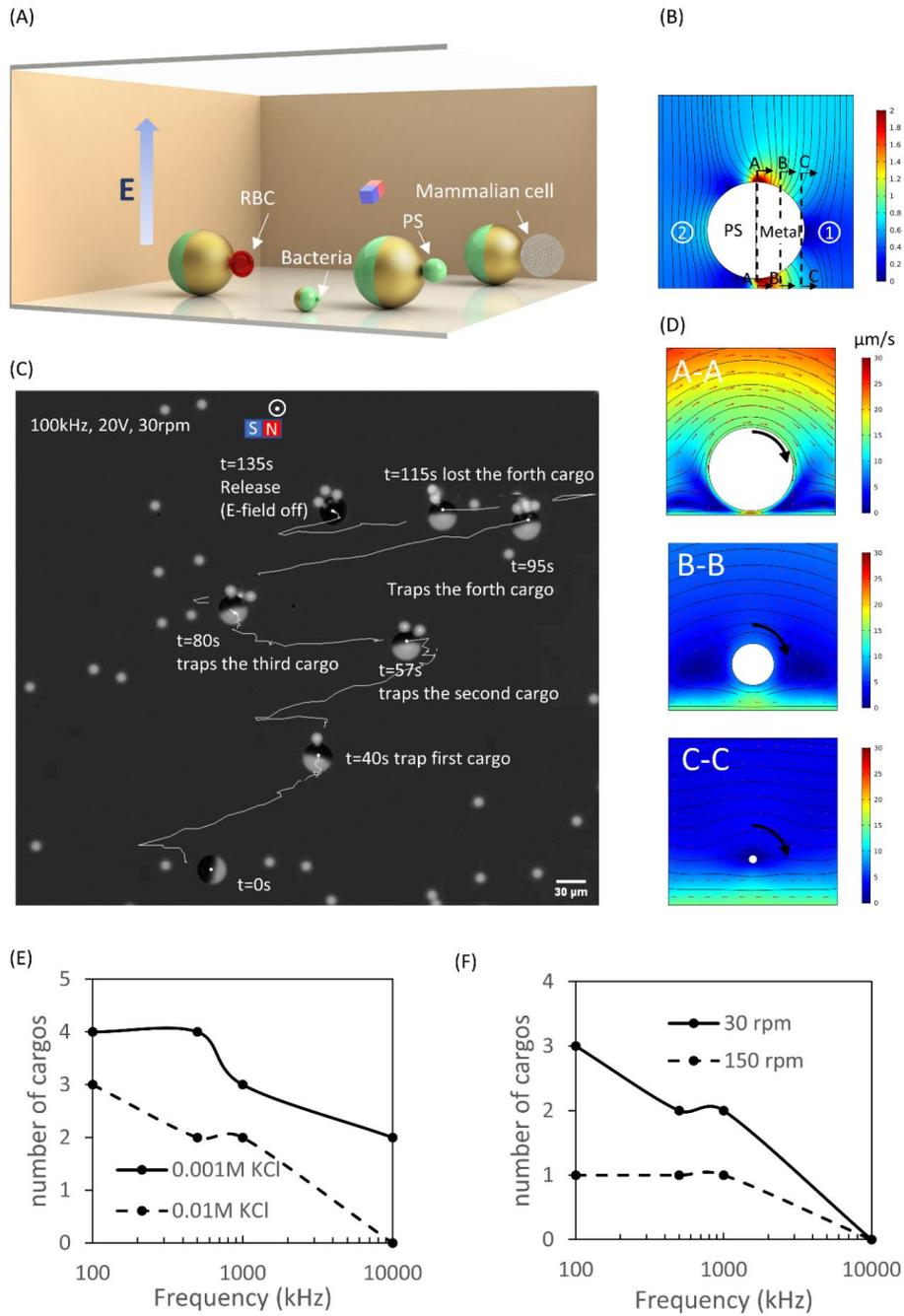

**Figure. 4. Selective cargo trapping and transport by a micromotor.** (A) Schematics of transport of a polystyrene particle (10µm), mammalian cell, and red blood cell (RBC) by a 27µm JP as well as transport of bacteria (*Rhodococcus*) by a 5µm JP within a conductive solution. (B) Numerically computed electric field intensity (normalized by the applied uniform electric field) and streamlines at the *x-y* plane. (C) Time-lapse and trajectory of a JP sequentially trapping three polystyrene particles. (D) Numerically computed velocity fields at A-A, B-B and C-C planes (marked in B). (E) The maximum number of trapped particles versus electric field frequency in 0.001M KCl and 0.01M KCl solutions, under 20V. (F) The maximum number of trapped particles versus electric field frequency for different magnetic field rotation speeds (30 rpm and 150 rpm) under 20V and solution conductivity of 1mS/cm, 0.01M KCl.

The micromotor also successfully trapped and transported biological cargos, e.g., bacteria, RBCs (see Video 4 in Supplementary materials) and cells (see Video 5 and 6 in Supplementary materials) in a wide range of solution conductivities, where trapping of each of these cargo types was based on their DEP characteristics (Table S2). As shown in Table S2, polystyrene spheres, mammalian cells (K562, time-lapse sequential images in Fig.5A, and Video 6 and 7 in Supplementary material), and RBCs (time-lapse sequential images in Fig.5B) were all trapped based on their nDEP response, and the trapping location was at the pole of the metallic-coated hemisphere (Fig 4B). However, *Rhodococcus* were pDEP in relatively low-conductivity solution (5% PBS, 95% 300mM sucrose, conductivity: 0.6-0.7mS/cm, pH: 7.31) at all frequencies, as well as in higher-conductivity solution (10% PBS, 90% 300mM sucrose, conductivity: 1-1.3mS/cm, pH: 7.31) at frequencies above 1MHz. Therefore, it was easier to trap *Rhodococcus* based on its pDEP characteristics at the polystyrene hemisphere pole (Fig.5C). It is worth noting at 100kHz, 20V in 20% PBS medium, the micromotor is able to trap and transport viable cell (Fig. S4.A, B and Video 7 in Supplementary Material) while avoiding its electroporation. This ability to transport cell intactly can be used as a single-cell drug delivery strategy as is demonstrated in Fig.S4C wherein a live cell was successfully transported at t=0s from a drug-free region to a region containing lipoplex with doxorubicin (DOX)-loaded liposomes (t=60s). The transported cell showed significant DOX uptake as compared to the non-transported control cells (t=120s).

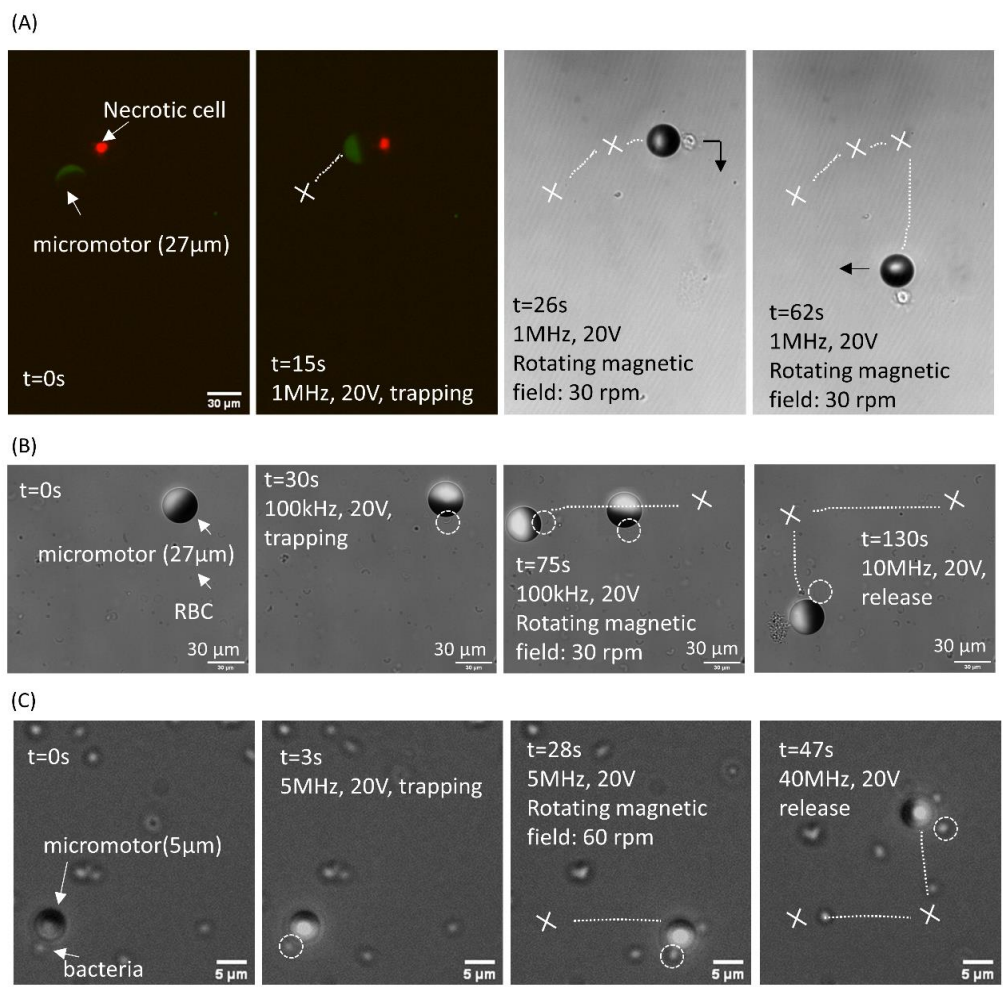

**Figure. 5. Sequential trapping and transport of various biological particles:** (A) Necrotic cell (K562) within 10% PBS and 90% 300mM sucrose solution (1-1.3mS/cm, pH: 7.31). The necrotic cell was stained with propidium iodide (red), which indicates that its membrane is not intact. (B) red blood cell (RBC) within 10% PBS and 90% 300mM sucrose solution (1-1.3mS/cm, pH:7.3). Refer to Video 4 and 5 in Supplementary materials and (C) Bacterial (*Rhodococcus*) cell within 1% PBS and 99% 300mM sucrose solution (0.1-0.15mS/cm, pH: 7.31).

## 2.5. Micromotor based Label free sensing of apoptotic, necrotic and viable cells

Apoptosis (known as programmed cell death) and necrosis (known as uncontrolled cell death) are considered as the two major cell death mechanism[49]. A cell undergoing apoptosis (can be triggered by doxorubicin DOX) will initiate a number of enzyme-dependent biochemical processes[49] which lead to the structural changes that can be reflected from the morphology (Fig. 6A). Annexin V staining (refer to Methods) is a common method for detecting apoptotic cells at their early stages of apoptosis[50] as shown in Fig. 6A for both 0.5 and 5μM DOX. Moreover, in Fig. S5, ~96±4% and ~98±2% of 0.5 and 5μM DOX treated cells, respectively, were stained with Annexin V. In 5μM DOX treated cell, the membrane blebbing can be observed, which indicates advanced apoptotic stage[51]. In contrast, both viable cells and necrotic cells cannot be stained by Annexin V. As shown in Fig. S5 ~12±5% of the viable cells (0 μM DOX) and 0% of the necrotic cells are stained by Annexin V. Cells undergo necrosis result in swelling and spillage of the contents of the cell into the extracellular space, which always cause inflammation response in the surrounding tissues[52] (Fig.6A Necrotic).

The micromotor system is able to identify necrotic cells induced by 3% $H_2O_2$ from viable cells and apoptotic cells based on their nDEP response at 1MHz within 10% PBS solution under an applied field of 20V (Fig.6 D, blue shaded region, see also Video 5) and as is verified in Fig.6B. The micromotor system is also able to identify early apoptotic cells (induced by 0.5μM DOX) from late apoptotic cells (induced by 5μM DOX), viable and necrotic cells as the DEP response of early apoptotic cells exhibits a cross over frequency (COF) response at ~200kHz within a 10% PBS solution, whereas viable, necrotic and late apoptotic cells exhibit nDEP response. The shift of COF of early apoptotic K562 cells might due to membrane capacitance increase[53, 54]. Hence, at 200kHz only the early apoptotic cells cannot be manipulated using DEP forces induced by the micromotor while the other types can (Fig.6B). The micromotor system is also able identify late apoptotic cells (induced by 5μM DOX) from early apoptotic cells (induced by 0.5μM DOX), viable and necrotic cells at ~100kHz within 10% PBS wherein the late apoptotic cells show stronger nDEP response at ~100kHz compared to the other cell type. The trajectory of the late apoptotic cells manipulated by JP (Fig.6E) shows a straight line due to the strong nDEP trapping force (~100% trapping in Fig.6B) rather than spiral line obtained for the other cell types due to the weaker nDEP force (~20-40% trapping in Fig.6B) imposed on the early apoptotic cells (induced by 0.5μM DOX) and viable cell which could not maintain its position at the center of the JP during its rolling motion (see also Video 6).

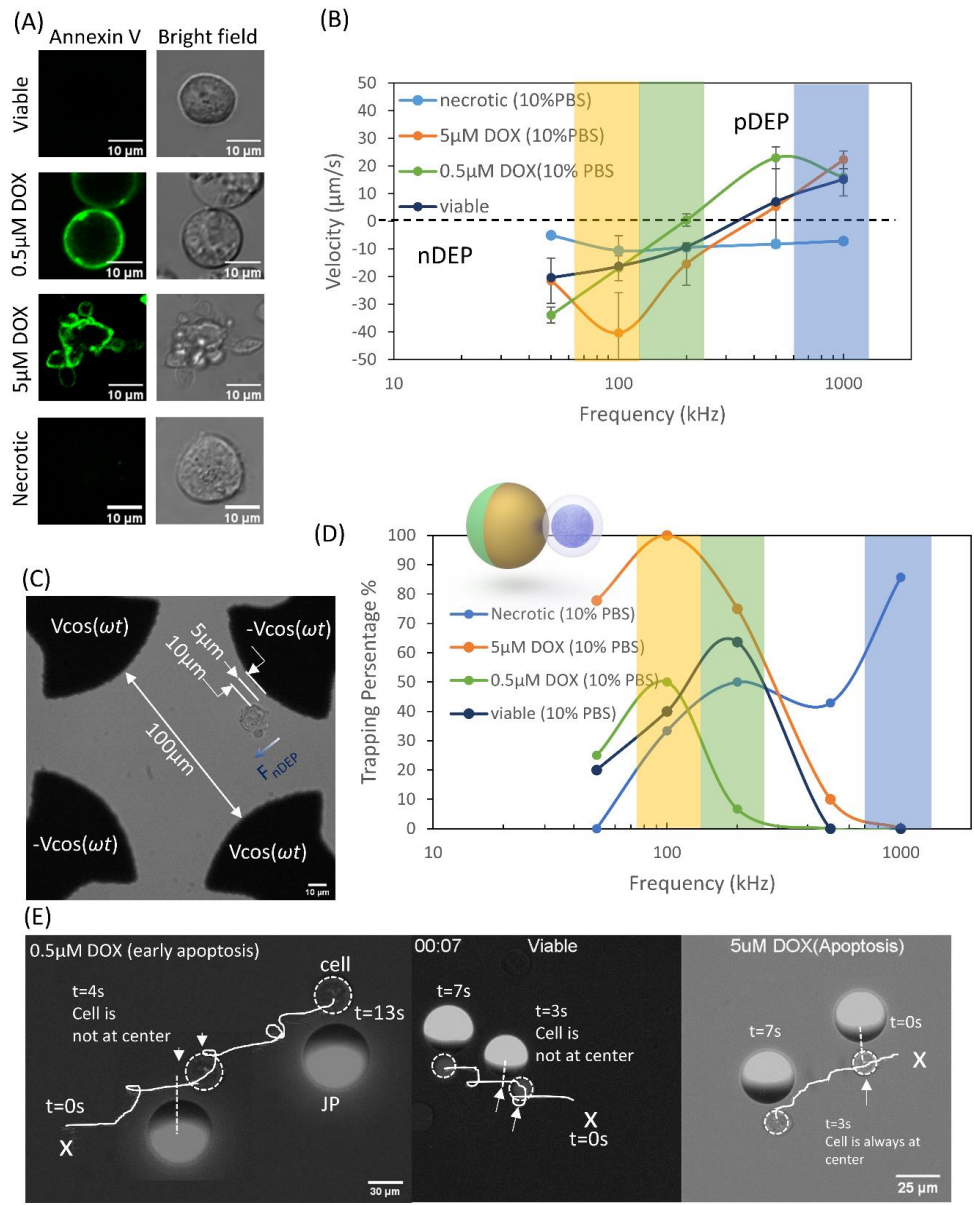

**Figure 6. Micromotor as a biosensor to identify and selectively manipulate apoptotic cells viable cells and necrotic among each other.** (A) the morphologies and Annexin V staining of viable, apoptotic (induced by 5μM and 0.5μM DOX) cells and necrotic cells (B) the velocity of cells measured between 5-10μm away from the nearby electrode for various medium and frequencies. The error bar is the standard deviation of 3 independent tests. (C) the dielectrophoretic response of cells is examined in a quadrupolar electrode set up. The velocity of a target cell is measured between 5-10μm away from the nearby electrode as the indication of the dielectrophoretic force. If the cell travels from the electrode to the center, the cell is experiencing negative dielectrophoretic force ($F_{nDEP}$); (D) The trapping rate of apoptotic cells, necrotic and viable cells in 10% PBS medium for various frequencies, 20V; (E) Trajectory of the trapped cells (viable, apoptotic induced by 5μM and 0.5μM DOX) manipulated by JP under 100kHz and 20V (see also Video 6 in Supplementary Material).

# 3. Conclusion and Discussion

Demonstrated herein was the use of a hybrid micromotor system synergistically propelled by both an external rotating magnetic field and applied electric field. More specifically, the electric field controlled the electro-orientation of the metallo-dielectric interface, which was also parallel to the direction of the pre-magnetized ferromagnetic Ni layer. Such electro-orientation resulted in a higher JP velocity due to the rotating magnetic field, as the magnetic dipole was forced to be in the same plane as that of the rotating magnetic field. Examination of this effect at varying solution conductivities, found it to be sustained up to 6mS/cm (0.05M KCl). At higher conductivities, the electro-orientation was not effective, due to the almost complete screening of the metallic coating. On the other hand, JP rolling induced by the rotating magnetic field, assisted in electrical-driven JP propulsion in an orthogonal direction. This resulted from the decreased friction of the JP with the substrate due to an induced lift force which led to its enhanced mobility and suppressed tendency to adsorb to the surface. Such combination of both magnetic and electric propulsion with the electric field-based mobility intensity and even direction (i.e., ICEP or sDEP), simply controlled by the frequency of the electric field, enabled better control of the JP trajectories. In contrast to previously reported synergistic effect between electric alignment and enhanced catalytic propulsion[55], our study reports not only the synergistic effect that the electric alignment has on the enhanced magnetic rolling induced velocity (Fig.2) but also vice-versa, i.e. the synergistic effect that magnetic rolling has on the electrical propulsion (Fig.3). Such complex two-way coupling (i.e. magnetic and electric field acting to enhance both magnetic rolling induced propulsion velocity as well as electrical propulsion) is novel and studied herein for the first time in a systematic manner involving varying key parameters, e.g. solution conductivity and electric field frequency.

The combination of magnetic and electric fields also enabled exploitation of DEP for cargo manipulation at relatively high solution conductivities where electrical propulsion (either ICEP or sDEP) is practically ineffective. Although oxygen-bubble propulsion mechanism of tubular microject[56] is proposed to address the ionic limitation of catalytic nanowire motors[57], this microject requires 5% hydrogen peroxide , which is toxic to most of the biological samples. In this paper, the DEP-based cargo manipulation is advantageous over other cargo-loading mechanisms due to its label-free nature, dynamic control over loading and release, polarizability-based selectivity (pDEP or nDEP response), and applicability for a broad range of organic and inorganic cargo sizes (1-17µm). In addition, the micromotor successfully trapped and transported various polystyrene spheres (3µm-15µm), and biological organisms, e.g., bacteria, RBCs and mammalian cells, and was able to differentiate between apoptotic, necrotic and viable mammalian cells with distinct DEP characteristics.

The ability of the micromotor to identify apoptotic and necrotic cells among viable cells based on their different dielectrophoretic response provides a way to perform label free sensing of cell status and its selective manipulation. At present moment, apoptosis is analyzed by flow cytometry[58], western blotting[59] or imaging methods[60]. There are also studies in which apoptotic cells are identified within a microfluidic platform due to the stiffness difference between them and viable cells[61]. Although micromotors have been widely studied in bio-sensing of antibodies[62], mRNA[63], and cancer cells[64], thus enabling real-time monitoring with single-cell/particle precision, our study is the first micromotor-based label free sensing of apoptotic cells. This ability opens new opportunities in study for a targeted single cell analysis. Apoptotic status can be analyzed or excluded for analysis in the single cell samples from drug discovery, developing of cell therapeutics and monitoring response in an immunotherapy. Moreover, dielectrophoresis can be used to reveal the membrane potential of biological changes of membrane[65] of a single cell or organelle (e.g. mitochondria)[66].

In our recent publication[67] using low-conductivity solutions for better electric propulsion, intact mammalian cells (K562) were not manipulable by the JP due to its pDEP characteristics, which led to its electroporation upon contact with localized high electric field regions on the JP surface. In the current study, the intact K562 was trapped and transported in higher-conductivity solutions (for example: 20% PBS, 80% 300mM sucrose, conductivity: 2.2-2.6mS/cm, pH: 7.31) via nDEP, with the trapping location situated at the local electric field minimum of the pole of the metallic coating at frequencies higher than 100kHz, which prevented cell electroporation[8] (Fig. S4 and Video 7 in supplementary material). Trapping and transporting live mammalians cell in conductive media opens new opportunities in drug delivery. For example, as shown here, mammalian cells can be targeted and transported to specific drug-enriched locations, precluding the need to load drugs into carriers[2]. Future developments can include cell modification to express viral toxin receptors, which will enable both on-demand drug delivery as well as viral toxin sensing, based on live cells. The ability to trap and transport live cells will also provide opportunities for cell fusion by bringing the targeted cell to contact another cell while applying an electric pulse[68].

It was found that beyond a certain solution conductivity, ~0.05M KCl (~6mS/cm), the induced–charge EDL screening rendered the electric field unable to provide sufficient dipole moment to align the JP's metallo-dielectric interface both parallel to the electric field and on the same plane as that of the rotating magnetic field. Moreover, the electric field gradients at the equator of the metallic surface might not provide adequate nDEP force for particle trapping. Therefore, ~6mS/cm might be the highest practical solution conductivity for this system, which is a bit lower than most physiological solutions, e.g., 1X PBS (13mS/cm), blood (10mS/cm), or mammalian cell culture media (~16mS/cm). However, by increasing the voltage to 75V and beyond, cells were successfully trapped at the equator of the metallic surface and were repelled at the dielectric surface (Video 8). At the same time, such high electric fields increased cell adsorption to the substrate and can lead to electrode burning. This limitation can be relaxed by using a lower voltage of 20V with a spherical JP of a larger (>27µm) diameter, which forms a stronger nDEP trap.

## 4. Experimental Section

*Experimental setup:* The experimental setup consisted of a microfluidic chamber formed by a 120µm-high spacer positioned between two parallel indium tin oxide (ITO)-coated glass slides (Fig. S7 in supplementary material). The bottom ITO-coated glass coverslip was coated with 15nm $SiO_2$ to prevent adsorption. Microbots (diameter: 5-27µm) were introduced into the chamber through the two 1mm-diameter holes. A block magnet was placed (Fig.1B) about 4cm away from the test chamber to induce an almost uniform (Fig. S2) magnetic field of ~5mT.

*Magnetic Janus particle fabrication:* Polystyrene particles (diameter: 27 µm, 15 µm) (Sigma Aldrich) in isopropanol (IPA) were pipetted onto a glass slide to form a monolayer upon solvent evaporation. The glass slide with particles was then coated with 15 nm Cr, followed by 50 nm Ni and 15 nm Au, as described by Pethig et al. [69] and Wu et al.[70]. The substrate was sonicated in deionized water (DIW) with 2% (v/v) Tween 20 (Sigma Aldrich), to release the JPs. The magnetic volumetric susceptibility of Au and Cr are negligible while that of Ni is 600. The magnetic steering of Janus particle is through magnetization of Ni.

*Cell culture and viability staining:* Human myelogenous leukemia K562 cells were grown at 37°C, 5% $CO_2$ in RPMI 1640 (Biological Industries), supplemented with 10% v/v heat-inactivated fetal bovine serum (FBS), 1% v/v penicillin-streptomycin (Biological Industries), and 2% v/v L-glutamine (Biological Industries). Cells were passaged every three days. Cells were incubated with 0.015mg/ml 6-carboxyfluorescein diacetate (CFDA) (37°C, 15 min). Necrotic cell death was induced by introducing 3% $H_2O_2$ into the cell suspension for 3 min. Apoptotic cells were prepared by adding 0.5 or 5µM doxorubicin and incubating for 18 hours. Intact, apoptotic and necrotic cells were centrifuged (600g, 5 min), three times, to replace the original solution with target solutions. Propidium iodide (PI) was added to the cell suspension in the target solution to achieve a final concentration of 3µg/ml. CFDA and PI fluorescent dyes were observed with lasers of wavelength 488 nm and 561 nm, respectively. Annexin V (Alexa Fluor 488) was purchased from ThermoFisher® where staining process followed the product protocol. Doxorubicin-loaded liposomes (Dox-NP) are purchased from Sigma Aldrich. The region close to the inlet with Dox-NP contained 20% Dox-NP and 80% 300mM sucrose in Fig. S4.

*Microscopy and image analysis:* Trapped and untrapped cargo was observed using a Nikon Eclipse Ti-E inverted microscope equipped with a Andor iXon-897 EMCCD camera. Images were captured using a x20 lens.

*Red blood cell (RBC) preparation.* Blood was received from Kaplan Hospital, Rehovot, Israel. The RBCs were harvested by centrifuging the blood samples at 500g for 10 min.

*Rhodococcus erythropolis preparation*: *Rhodococcus erythropolis* ATCC 4277 were cultured in a lysogeny broth (LB) agar plate (30°C, 48 h). Bacterial colonies were transferred to PBS. Cells were washed 3 times with target solution. Before the experiment, 3µg/ml PI and 0.1% Tween 20 were added to the solution and cells were incubated for 5 min, at room temperature. All reagents were purchased from Sigma-Aldrich.

*Polystyrene particle solution:* Particles were rinsed three times with target solutions, to which Tween 20 (0.1% (v/v)) was added to minimize adhesion to the ITO substrate before being mixed with the microbots.

*Quadrupolar electrode array experimental setup (Fig. 6C):* 10μL of a solution comprised of targeted cell types (viable, 0.5μM DOX/5μM DOX treated cells, or 3%H2O2 treated cells) with a concentration of ~$10^5$ cells/ml. within 10% PBS and 90% 300mM sucrose were introduced into the closed Silicone reservoir (120μm height) above the quadrupolar electrode. During the DEP experiment, only one or few cells were placed within electrode array to prevent cell-to-cell electrical interaction. Various AC field frequencies with a sinusoidal wave form were applied using function generator (33250A, Agilent). In order to minimize the side effects (e.g. electrolysis, Joule heating) of the cells, a sufficiently low-amplitude AC field ($4V_{pp}$) was applied. The motion of the cell was recorded using Andor Neo sCMOS camera (10 frames per second) attached to a Nikon Ti inverted epi-fluorescent microscope with 20 x objective lens and further analyzed by ImageJ.

*Numerical simulations:* The numerical simulation of the electric field and electric streamlines (Fig.2 (F, G)) was performed in COMSOL™ 5.3. A simple 2D geometry, consisting of a rectangular channel, 25 μm height and 20 μm width, with a 10 μm-diameter circle placed 300nm above the substrate, was used to model the experimental setup. Since the EDLs are thin relative to the radius of the particle $\left(\frac{\lambda}{a} \ll 1\right)$, within the electrolyte, we can solve the Laplace equation for the electric potential, $\phi$, in conjunction with the following boundary condition at the metallic side of the JP

$$\sigma \frac{\partial \phi}{\partial n} = i\omega C_{DL}(\phi - V_{floating}), \quad (1)$$

which describes the oscillatory Ohmic charging of the induced EDL, wherein $V_{floating}$ is the floating potential of the metallic hemisphere of the microbot, *n* is the coordinate in the direction of the normal to the microbot surface, and $C_{DL}$ represents the capacitance per unit area of the EDL and can be estimated from the Debye-Huckel theory as $C_{DL} \sim \varepsilon/\lambda$. In addition, a floating boundary condition[44] was applied on the metallic hemisphere so as to obey total zero charge. An insulation boundary condition was applied on the dielectric hemisphere of the JP, a voltage of 6.25 V was applied at the lower substrate (*y*=0), the upper wall was grounded, and the edges of the channel were given an insulating boundary condition. In the numerical simulation for the fluid velocity and pressure induced around the micromotor upon its magnetic rolling (Fig.3G and Fig.4B, D), boundary conditions are listed in Table S1.

## Acknowledgments

G.Y. acknowledges support from the Israel Science Foundation (ISF) (1934/20). Y.W. acknowledges support from the Technion-Guangdong project for postdoctoral fellowship. Fabrication of the chip was made possible through the financial and technical support of the Russell Berrie Nanotechnology Institute and the Micro-Nano Fabrication Unit. We thank the lab of Prof. Aaron Ciechanover for providing the K562 cell line. We thank Prof. Kashi for providing *Rodococcus erythropolis ATCC 4277*. We thank Kaplan Hospital for providing the blood sample. We also thank Eyal Abraham and Shlomo Wais for constructing the joysticks for manipulating the magnetic field controlled by Arduino Nano.

## Conflict of Interest

The authors declare no conflict of interest.

## Keywords

Hybrid micromotor, Active colloids, Dielectrophoresis, Cargo transport